# HRnV-Calc: A software package for heart rate n-variability and heart rate variability analysis

Chenglin Niu[1,#], Dagang Guo[1,#], Marcus Eng Hock Ong[1,2], Zhi Xiong Koh[1,2], Andrew Fu Wah Ho[1,2], Zhiping Lin[3], Chengyu Liu[4], Gari D. Clifford[5,6], Nan Liu[1,7,8]*

[1] Duke-NUS Medical School, National University of Singapore, Singapore, Singapore
[2] Department of Emergency Medicine, Singapore General Hospital, Singapore, Singapore
[3] School of Electrical and Electronic Engineering, Nanyang Technological University, Singapore
[4] School of Instrument Science and Engineering, Southeast University, Nanjing, China
[5] Department of Biomedical Informatics, Emory University, Atlanta, GA, United States of America
[6] Department of Biomedical Engineering, Georgia Institute of Technology, Atlanta, GA, United States of America
[7] SingHealth AI Health Program, Singapore Health Services, Singapore, Singapore
[8] Institute of Data Science, National University of Singapore, Singapore, Singapore

[#] Contributed equally
* Corresponding Author:
Nan Liu
Programme in Health Services and Systems Research
Duke-NUS Medical School
8 College Road
Singapore 169857
Singapore
Phone: +65 6601 6503
Email: liu.nan@duke-nus.edu.sg

## Abstract

*Objective:* Heart rate variability (HRV) has been proven to be an important indicator of physiological status for numerous applications. Despite the progress and active developments made in HRV metric research over the last few decades, the representation of the heartbeat sequence upon which HRV is based has received relatively little attention. The recently introduced heart rate n-variability (HRnV) offers an alternative to R-to-R peak interval representations which complements conventional HRV analysis by considering HRV behavior on varying scales. Although HRnV has been shown to improve triage in pilot studies, there is currently no open and standard software to support future research of HRnV and its broader clinical applications. We aimed to develop an open, reliable, and easy to use software package implementing HRnV for further research and improvements of HRnV. This package has been designed to facilitate collaborative investigations between clinicians and researchers to study HRnV in various contexts and applications.

*Approach:* We developed an open-source software, HRnV-Calc, based on the PhysioNet Cardiovascular Signal Toolbox (PCST), which features comprehensive graphical user interfaces (GUIs) for HRnV and HRV analysis.

*Main results:* While preserving the core functionalities and performance of PCST, HRnV-Calc enables step-by-step manual inspection and configuration of HRV and HRnV analysis, so that results can be debugged, easily interpreted, and integrated to downstream applications.

*Significance:* The open-source HRnV-Calc software, an accessible and standardized HRV and HRnV analysis platform, enhances the scope of HRV assessment and is designed to assist in future improvements and applications of HRnV and related research.

**HRnV-Calc Software**: https://github.com/nliulab/HRnV-Calc

# 1 Introduction

Variation of the time interval between a consistent point in time of each heartbeat (generally related to ventricular contraction), known as heart rate variability (HRV) [1], has been proven by numerous studies to be a useful indicator of physiological status [1-3]. Thanks to its non-invasive nature and strong connection to the autonomic nervous system (ANS) [1, 3], HRV has been adopted to study a wide range of diseases and clinical conditions, which include myocardial infarction [4-6], sudden cardiac death [5, 7], diabetes [8-10], renal failure [11, 12], sepsis [13, 14], and cancer [15-17]. In addition, the emergence of wearable devices with heart monitoring capabilities has also allowed researchers to study the above-mentioned medical conditions in real-world settings [18-20], as well as in non-clinical applications such as sports [21], stress [22], and sleep monitoring [23].

A variety of HRV metrics can be derived from the heartbeat time sequence known as the inter-beat interval (IBI), or the R-to-R peak interval (RR interval [RRI]). Such sequences are often extracted from biomedical signals such as electrocardiograms (ECG) and photoplethysmograms (PPG). It is believed that a decrease of complexity in HRV is associated with an increase in both morbidity and mortality [1, 3-5]. To qualify and quantify the complexity, various conventional HRV metrics in linear and non-linear domains [24] have been established to reflect the dynamic of HRV [1, 25]. However, the exact mechanism regulating HRV is not well understood [1-3]. Recent developments in HRV have been mainly focused on non-linear metrics such as variants of approximate entropy (ApEn) and sample entropy (SampEn) [26-28].

Despite progress in HRV metrics research, the representation of RRI upon which HRV is based has rarely been examined. Cysarz et al. [29] proposed a binary symbolization of RRI, which combined with ApEn provides new information about the normal heart period regularity. The multiscale entropy (MSE) metrics [30] calculate SampEn on multiscale coarse-gained series derived from RRI to reflect the nonlinear behavior of the heart on multiple time scales. To generalize the averaging multiscale approach, Liu et al. [31] proposed heart rate n-variability (HRnV) that utilizes sliding and stridden summation windows over RRI to obtain new RRI-like intervals denoted $RR_nI$ and $RR_nI_m$. Using these novel RRI representations, new HRnV metrics can be calculated with conventional HRV analysis metrics, providing an entire family of new metrics, and potentially additional insights into the dynamics and long-term dependencies of the original RRI. For example, research has shown that HRnV improves the accuracy of triage for patients with chest pain [31] and sepsis [32].

For the reliable and repeatable implementation of HRnV in both clinical and research settings, an open-source and standardized software platform is necessary to facilitate further research on HRnV and its possible variations. There is an abundance of HRV software tools for commercial and non-commercial use, including Kubios HRV [33], ECGlab [34], ARTiiFACT [35], rHRV [36], and RR-APET [37]. However, none of these tools is suitable for incorporating HRnV analysis. Moreover, with the exception of two open-source toolboxes, none of the available tools provide equivalent results, making comparisons between research impossible [38]. Since HRnV shares some common processing methods with conventional HRV analysis [31], it is natural to develop a HRnV platform based on existing benchmarked software. We therefore developed an open-source HRnV software, HRnV-Calc, based on the PhysioNet Cardiovascular Signal Toolbox (PCST) [38]. Compared to other HRV freeware, the PCST is standardized and well-documented. More importantly, the PCST is an open-source HRV software suite which has gone through rigorous testing and benchmarking in both technical and clinical settings. Based on the fully functional HRV command-line code provided by the PCST, HRnV-Calc has integrated graphical user interfaces (GUIs) that enable manual inspection and correction of RRI extraction from ECG signals, flexible configuration, and batch-processing. Its inherent functions support the analysis of both HRnV and conventional HRV metrics. Therefore, HRnV-Calc facilitates new methodological developments, but also provide clinicians and researchers with standardized HRnV and HRV analyses.

## 2 Computational Methods

In HRnV-Calc (https://github.com/nliulab/HRnV-Calc), the main components of signal processing and HRV analysis are built upon the PCST. This section provides an overview of the methods implemented in HRnV-Calc as well as the HRnV method it integrates. Table 1 contains a complete list of metrics provided by HRnV-Calc.

### 2.1 QRS Detection

If inputs to HRnV-Calc are raw ECG waveforms, the software will first attempt to detect QRS features. The QRS detector used in HRnV-Calc is *jqrs* [39, 40], as the same implemented in the PCST [38]. Upon finalization of the R peak detection results, the ECG data will be converted to RRI sequence for downstream analysis.

### 2.2 Preprocessing of RRI

Both user-provided RRI and ECG-derived RRI will undergo a preprocessing phase to determine whether the input data contains non-sinus intervals or outliers [38]. A non-sinus interval is detected by measuring whether the current interval exceeds the median of five adjacent intervals by a user-prespecified threshold, which has a default value of 20% [38]. Users may choose to remove such intervals from the data or to interpolate new intervals using cubic spline, piecewise cubic Hermite interpolating polynomial (PCHIP), or linear interpolation.

## 2.3 HRnV Method

The HRnV method [31] for alternative RRI representation is a unique and the main feature implemented in HRnV-Calc. HRnV utilizes sliding and stridden summation windows on the original RRI, resulting in new $RR_nI$ and $RR_nI_m$ intervals, which can then be fed into conventional HRV analysis to calculate corresponding $HR_nV$ and $HR_nV_m$ metrics. For clarification, the term 'HRnV' refers to the name of the method (i.e., heart rate n-variability), while $HR_nV$ and $HR_nV_m$ refer to the analysis and derived metrics based on $RR_nI$ and $RR_nI_m$ intervals, respectively.

HRnV analysis requires specifications of two parameters: the summation parameter $n$ and the stride parameter $m$. Originally, $RR_nI$ and $RR_nI_m$ are two distinct representations derived from RRI [31]. $RR_nI$ is generated by non-overlapping summation windows, while $RR_nI_m$ is created from overlapping ones. Mathematically, $RR_nI$ can be treated as a special case of $RR_nI_m$ (i.e., $m = n$, or $RR_nI_n$). To simplify the terminology, in the remainder of this paper we only use $RR_nI_m$ and $HR_nV_m$ to represent the intervals and analysis provided by HRnV, unless specifically stated.

Both $n$ and $m$ can take any positive integer values (i.e., $n, m >= 1$) given $m <= n$. To describe the process of HRnV, consider a series of clean input RRI as $X_i$ ($i = 1,2,3, \ldots, N$) of length $N$. With specified parameters $n$ and $m$, a new series of $RR_nI_m$ intervals, $Y_i$ ($i = 1,2,3, \ldots, M$) of length $M$ ($M <= N$) can be expressed as:

$$Y_i = \sum_{j=1}^{n} X_{(i-1)*m+j}, (i = 1,2,3, \ldots, M)$$

The length of the new intervals, $M$, is given by $M = \left\lfloor \frac{N-n+1}{m} \right\rfloor$, where $\lfloor \cdot \rfloor$ represents the floor function. Essentially, the original RRI is equivalent to $RR_1I$ intervals.

The $RR_nI_m$ intervals capture and amplify the dependencies and dynamics within the original RRI. The parameter $n$ determines how HRnV smooths and amplifies the original RRI, while the parameter $m$ indicates how much dynamics it attempts to capture.

All conventional HRV metrics mentioned in the subsequent subsections can be calculated from $RR_nI_m$ intervals. These metrics, also known as $HR_nV_m$ metrics (e.g., $HR_2V_1\_SampEn$), provide potential new insights of the original RRI, especially in the frequency and nonlinear domains. A single $HR_nV_m$ analysis with specified $n$ and $m$ is an HRV analysis performed on the corresponding $RR_nI_m$ interval. Consequently, the conventional HRV analysis is equivalent to the $HR_1V$ analysis.

## 2.4 Time-domain Methods
HRnV-Calc provides the majority of commonly used time-domain metrics [25]. Statistically descriptive metrics such as the mean, standard deviation (SDRR), skewness, kurtosis, and the root mean square of successive differences (RMSSD) are calculated based on functions provided in the PCST. Additionally, HRnV-Calc calculates the mean and standard deviation of heart rates (HR) as well as the HRV triangular index [25].

## 2.5 Frequency-domain Methods
In frequency-domain analysis, the power spectral density (PSD) estimation is first derived from input intervals. Methods for such an estimation provided in HRnV-Calc include the default Lomb-Scargle periodogram [38, 41, 42], Welch PSD estimate, discrete fast Fourier transform (FFT), and Burg PSD estimate.

The estimates of PSD are divided into three bands: very low frequency (VLF), low frequency (LF), and high frequency (HF). Default limits of these three bands are 0 to 0.04 Hz for VLF, 0.04 to 0.15 Hz for LF, and 0.15 to 0.4 Hz for HF.

The frequency-domain HRV metrics computed in HRnV-Calc include the peak frequency for each of the three specified bands, absolute and relative powers of each band, normalized power of LF and HF, LF/HF power ratio, and the total power of PSD.

## 2.6 Nonlinear Methods
The nonlinear methods included in HRnV-Calc are the Poincaré plot [43], approximate entropy, sample entropy [26], and detrended fluctuation analysis (DFA) [44].

Poincaré plot is a graphical examination of the correlation within the input intervals [43]. The width ($SD1$) and the length ($SD2$) of the eclipse fitted on the plot capture short-term and long-term dependencies of RRI, respectively.

ApEn and SampEn measure the degree of irregularities in the intervals. The embedding $m$ and tolerance $r$ must be specified to calculate such entropies. By default, HRnV-Calc sets $m = 2$ and $r = 0.15$ SDRR [38].

DFA captures short-term and long-term fluctuations by fitting two slopes ($\alpha_1$ and $\alpha_2$) in the log-log plot [44], which is a graphical representation of correlations within RRI for different time scales. In HRnV-Calc, $\alpha_1$ is fitted in the plot within 4 to 16 beats, while $\alpha_2$ is fitted for beats no less than 16 [38].

## 3 Software Description

The main features of HRnV-Calc are HRnV metrics calculations and GUIs for both HRV and HRnV analysis, which are built upon the core HRV analysis command-line code provided by the PCST. Therefore, installation of the PCST is required before using HRnV-Calc, which is available on GitHub [45].

In this section, we will introduce the functionalities built-in HRnV-Calc. The ECG file 'Demo_NSR16786.txt', extracted from MIT-BIH Normal Sinus Dataset [46, 47] is used as the demonstrative input to HRnV-Calc. Details of the demo input file can be found in section 5.

HRnV-Calc is primarily operated using its step-by-step GUIs, which include four main interfaces: (1) Data Loader, (2) QRS Detection & Edits viewer, (3) $HR_nV_m$ Setting viewer, and (4) $HR_nV_m$ Results Display. Each of these interfaces will be presented one at a time for every step of HRnV and HRV analysis.

### 3.1 Data Loader

As shown in Figure 1(a), the initial GUI of HRnV-Calc is Data Loader, which provides basic settings for users to begin HRV/HRnV analysis, such as the data type of the input and the sampling rate of the signal. Users can choose to perform analysis on a single file or multiple files as batch-processing. It is noteworthy that the current version of HRnV-Calc only supports batch processing on RRI inputs, which do not require manual QRS inspection to complete the HRV/HRnV analysis.

The 'Fetal ECG' section allows users to select QRS detection profiles based on the type of input ECG, since a fetal ECG requires completely different QRS detection settings from ECG captured from an individual. The two profiles currently provided in HRnV-Calc are based on the profiles given in the PCST [38].

HRnV-Calc provides three options for input data type: Raw ECG (*.txt, *.csv), IBI (*.txt, *.csv), and ECG PC (peak corrections) with peak annotations (*.csv) saved by HRnV-Calc. For ECG files, currently HRnV-Calc only supports single channel signals. Both single channel ECG and RRI data should be stored in a single column or row in the supported file formats. Examples of the data formats currently supported by HRnV-Calc can also be found in the sample data inputs provided in the supplementary.

By default, HRnV-Calc will use the full name of input files as the record ID to display and store downstream HRV/HRnV analysis results. Users may choose to specify the prefix and postfix of the file name for ID extraction. Once all settings of the input are set, the 'Next' button will bring up a confirmation window (Figure 1(b)) to remind users of the initial settings before proceeding to the subsequent step.

### 3.2 QRS Detection and Edits (QDE) Viewer

The QDE viewer (Figure 2) is designed to configure and inspect QRS detection on ECG inputs. All settings and tools for QRS detection and inspection can be found in the setting section at the top. The 'Signal Type' option allows users to choose whether the full ECG or a segment of it should be analyzed.

Once the segmentation of ECG is finalized, users may proceed to perform R peak detection on the selected segment. Additionally, baseline drift can be removed before QRS detection. Users may also choose to perform local checkup on the ECG input to modify the R peaks to local maximum or minimum of the signal. Details of the local checkup can be found in the supplementary. The processed ECG and R peak positions (marked as red dots in Figure 2) will be plotted at the bottom of the QDE viewer when QRS detection is initiated.

Inspection of QRS detection results can be done in the Denoised ECG Plot in QDE viewer as shown in Figure 2. If manual correction of R peaks is required, first remove the incorrect R peaks by clicking on the 'Remove Peak' button, and then add new annotations using 'Add Peak' button.

### 3.3 $HR_nV_m$ Setting

$HR_nV_m$ analyses are configured in the $HR_nV_m$ Setting viewer (Figure 3). In $HR_nV_m$ configuration, users may choose to perform a single $HR_nV_m$ analysis by choosing the option 'Single' and specifying the desired $n$ and $m$ values. When choosing the option '$m = n$', HRnV-Calc will perform $HR_nV$ (i.e., $HR_nV_n$) analysis on the input depending on the specified $n$. By default, HRnV-Calc will perform the conventional HRV analysis with $n = 1$ and $m = 1$.

The option 'All' lets users perform all $HR_nV_m$ analyses with $n$ and $m$ no greater than the specified $n$. For example, if $n$ is set to be 2, HRnV-Calc will conduct $HR_1V$ (i.e., conventional

HRV), $HR_2V_1$, and $HR_2V$ analyses on the input signal altogether. The default value of $n$ for this option is 1, indicating only the conventional HRV analysis to be performed.

The 'Ectopic Beats' section allows users to specify the threshold for a beat to be considered as an outlier and to select how outliers should be processed. It should be noted that the detection and processing of ectopic beats will only be conducted on the original RRI. All $RR_nI_m$ intervals will be generated from the processed RRI intervals without further processing. In the 'Frequency Domain' section, users may choose one of the four PSD estimation methods provided in the PCST [38].

### 3.4 $HR_nV_m$ Results Display

HRnV-Calc will display (Figure 4) the results of a single $HR_nV_m$ analysis (e.g., $HR_2V_1$) in the $HR_nV_m$ Results Display window. Note the display window will not be activated if users choose to perform multiple $HR_nV_m$ analyses (by selecting the 'All' option in $HR_nV_m$ Setting) or to conduct batch processing on multiple input files.

As shown in Figure 4, the $HR_nV_m$ Results Display viewer provides a comprehensive overview of the $HR_nV_m$ analysis. If the conventional HRV analysis is performed, the 'IBI Statistics' section provides an overview of the abnormal beats presented in the original RRI and the percentage of clean beats in the entire input. For $HR_nV_m$ analyses other than conventional HRV, the 'IBI Statistics' section will only display the number of beats in the corresponding $RR_nI_m$ intervals, as preprocessing is only performed on the original RRI before converting to $RR_nI_m$ intervals.

## 4 HRnV-Calc Validation

To ascertain the quality and credibility of HRnV-Calc, we compared it with the PCST. Since HRnV-Calc is built upon the PCST, both QRS detection and HRV metrics calculation in HRnV-Calc should be compared with the PCST to ensure that core functionalities of the PCST are properly inherited.

In this section, we validate the QRS detection, HRV, and $HR_nV_m$ calculations from HRnV-Calc with the PCST. The MIT-BIH Normal Sinus Dataset [46, 47] was used as the input to test QRS detection. For each patient in the dataset, a five-minute ECG recording was randomly selected. A total of 18 five-minute ECG recordings were fed into both software under identical analysis configurations.

First, we compared the QRS detection results obtained by the PCST and HRnV-Calc. The $l_1$ distance, $d_{l_1}$, was used to measure differences between R peak annotations generated by the two software. The average distance, $\overline{d_{l_1}}$, of all test inputs was 0. Therefore, QRS detection from HRnV-Calc yielded identical results to those obtained using the PCST. Details of the QRS annotation comparison can be found in the supplementary material.

Subsequently, we compared HRV metrics calculations between the two software packages using RRIs derived from ECGs. To examine the differences between the calculated HRV metrics, we used the relative error $\epsilon_x$ of metric $X$, defined as $\epsilon_x = \frac{|X_H - X_P|}{|X_P| + 10^{-8}}$, where $X_H$ and $X_P$ represented the same metric $X$ generated by HRnV-Calc and the PCST for one patient, respectively. For every common metric calculated by the PCST and HRnV-Calc, we calculated the average relative error of the metric among all 18 patients. As a result, all average errors for all commonly shared HRV metrics were negligible (i.e., smaller than $10^{-4}$), indicating that HRnV-Calc performed HRV analysis identically to PCST.

Furthermore, we compare the $HR_nV_m$ metrics calculated by the PCST and HRnV-Calc. Since conventional HRV software packages do not have built-in functions for HRnV analysis, we exported the $RR_nI_m$ intervals from HRnV-Calc and provided them as normal inputs to the PCST. Such analyses, however, are not recommended as most HRV software packages are not designed to analyze $RR_nI_m$ intervals. To simplify the comparison procedure, we only performed $HR_2V_1$ and $HR_2V$ analyses for demonstration.

Thanks to the open-source command-line code provided by the PCST, we were able to configure the PCST to properly process the $RR_nI_m$ intervals constructed from the 18 RRIs, which were derived from ECGs as previously described. As a result, the averaged relative error for every $HR_2V_1$ or $HR_2V$ metric calculated by the two software was negligible (i.e., smaller than $10^{-4}$), demonstrating identical results.

We therefore conclude that HRnV-Calc performs identical to the PCST in terms of QRS detection and HRV metrics calculation under the same configuration. The PCST can also be modified to process $RR_nI_m$ intervals identical to HRnV-Calc. Details of the exact HRV and HRnV metrics calculated by both software can be found in the supplementary materials. For more comparisons between the PCST and other HRV analysis software, see Vest et al. [38].

## 5 Demonstration Code

To provide a demonstration run of the software, a ten-minute ECG signal (sampling rate of 128Hz) was randomly segmented from patient #16786 in the MIT-BIH Normal Sinus Database [46, 48](found in file 'Demo_NSR16786.txt'). A five-minute segment (the second half of the signal) was then selected in the QDE viewer. Baseline drift was removed from the segment prior to QRS detection, after which no manual R peak correction was made. HRV metrics calculated from the selected segment are shown in Figure 4.

# 6 Discussion

In this paper, we introduce HRnV-Calc, a software package based on the PCST, that allows for a standardized and reliable implementation of HRnV and HRV analyses in both research and clinical settings. While preserving core functionalities and performance of the PCST, HRnV-Calc provides intuitive GUIs for manual inspection and configuration at each step of HRnV and HRV analyses, in addition to the fully automated processing provided in the PCST [38]. Such inspection and configuration are vital to achieving scientifically valid and universally comparable HRnV and HRV results, particularly for clinical applications, while also increasing the credibility and interpretability of analysis results. The batch-processing function enhances processing efficiency when analyzing multiple RRI records simultaneously. More importantly, the standardized and well-benchmarked methods included in the PCST and HRnV-Calc greatly facilitate the establishment of transparency and consensus on method configurations among HRV researchers, thereby reducing misunderstandings and sometimes conflicting results between studies using different HRV software packages [38].

HRnV is a novel HRV analysis method unique to HRnV-Calc. The $RR_nI_m$ intervals generated by HRnV can be viewed as natural extensions of the RRI representations. By incorporating $RR_nI_m$ intervals into existing HRV analysis, HRnV provides new insights into the dynamic and long-term dependencies of the original RRI. In recent studies, HRnV has shown promising results in boosting the accuracy of triage for patients with chest pain [31] and sepsis [32], as compared to triage based on conventional HRV. HRnV can be used to complement conventional HRV analysis to provide additional insight. Using HRnV-Calc, HRnV can now be easily incorporated into any conventional HRV study by a few configurations in the GUI, regardless of users' programming knowledge.

Although HRnV has been shown to provide new insights, its physiological interpretation, and the exact mechanism it captures new information remain unclear. Moreover, the abundance of metrics generated by HRnV provides may also raise concerns about redundancy and intercorrelation within the data [31]. Based on previous studies utilizing HRnV, correlation analysis may be necessary to provide more informative HRnV results [31, 32]. However, the standard data processing method for HRnV metrics remains unclear and requires further research of HRnV in various contexts and applications. An alternative way to uncover insights in HRnV metrics may be trend analysis over multiple $m$ or $n$ parameters, which is similar to the analysis proposed in MSE [30]. In addition, little consensus exists regarding the selection of HRnV parameters $n$ and $m$, which seem to capture more long-term dependencies with larger values, at the cost of a significant reduction in signal length. The introduction of HRnV-Calc aims to address these concerns about HRnV by providing an easily accessible and user-friendly software platform for clinicians and researchers to utilize and study HRnV in their own research. In doing so, we hope HRnV-Calc will facilitate future applications and improvements to HRnV, while potentially supporting further research into biomedical signal representations.

The HRnV-Calc tool remains in active development. There are several limitations in the software. First, the current version of HRnV-Calc was primarily designed for short ECG/RRI recordings (between five minutes and one hour). The software has not been optimized for analyzing long signals, such as 24-hour ECG recordings. Second, although HRnV-Calc supports batch-processing of multiple RRI inputs, which do not require manual inspections to proceed to HRV or HRnV analysis, the step-by-step workflow in HRnV-Calc may prevent the analysis of large quantities of raw ECG data in certain use cases. On the other hand, for general clinical use, the GUIs and manual operations in HRnV-Calc provide users with opportunities to fully control the processing of raw signals, thereby ascertaining the quality of RRI extraction from raw ECG data for subsequent HRnV and HRV analysis. Moreover, users with sufficient programming skills can use the source code provided in HRnV-Calc and the PCST to conduct fully automated and customized HRnV analysis.

## 7 Conclusions

HRnV is a useful method for providing new insights into biomedical signals. Its software implementation, HRnV-Calc, is a GUI-based, reliable software package for HRnV and HRV analyses. HRnV-Calc is an open-source project, and we encourage collaboration and improvements from researchers and programmers within the larger HRV research community to bring new features and experiences into future editions of the software.

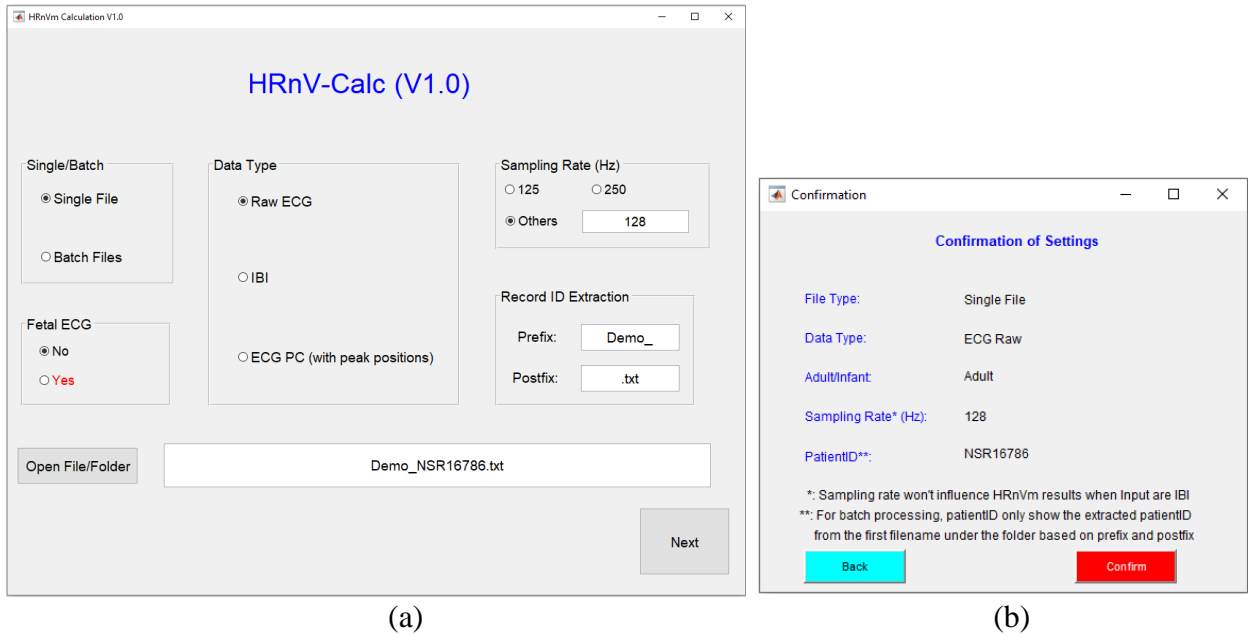

Figure 1 – Data Loader (a) and Setting Confirmation (b)

Data Loader (a) allows users to specific basic settings of the input data before the analysis begins. Once the settings are finalized, the 'Next' button will bring up the confirmation window (b) before proceeding to the next step of analysis.

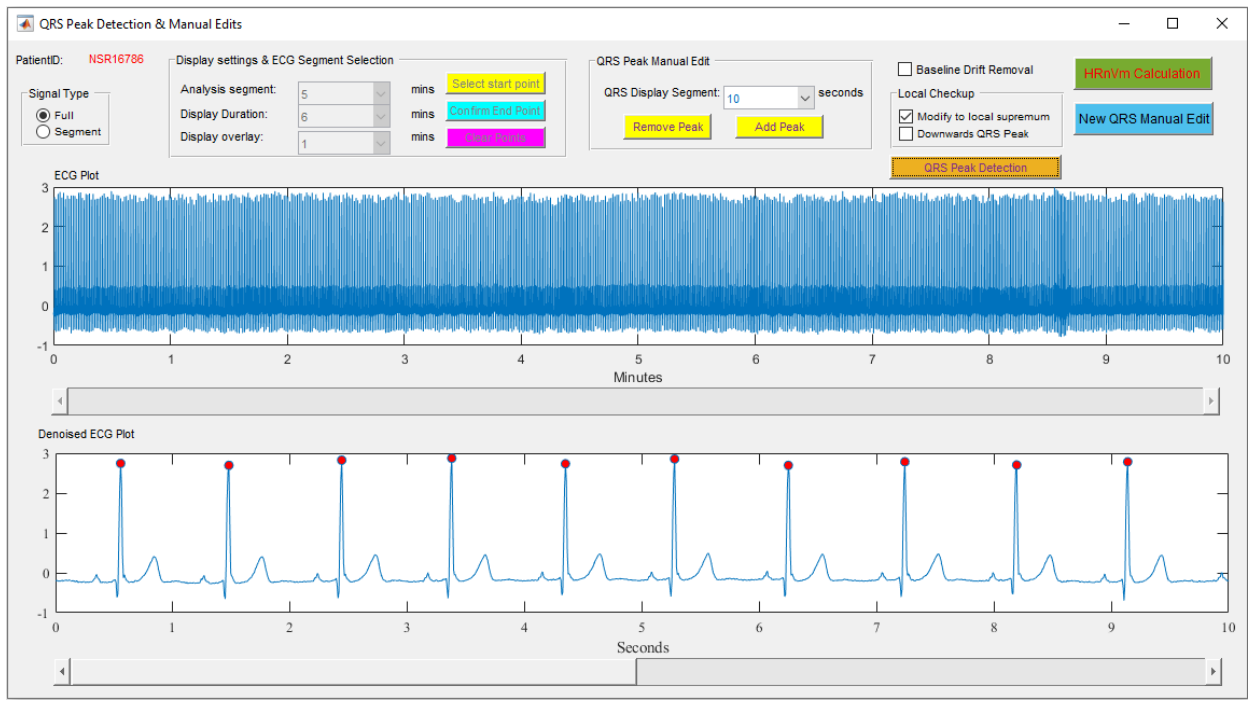

Figure 2 – The QRS Detection and Edits (QDE) Viewer

The QDE Viewer allows users to configure and inspect QRS detection on ECG input. ECG display and segmentation settings along with QRS preprocessing settings are located at the top of the viewer. R peak annotations can be checked and modified in the Denoised ECG Plot at the bottom of the viewer.

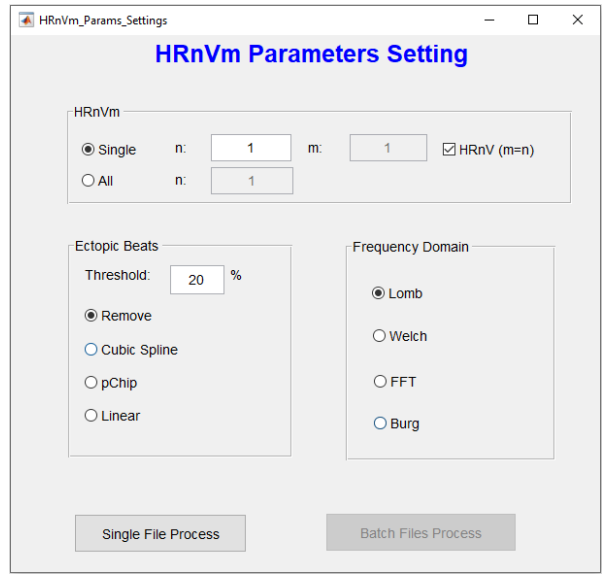

Figure 3 – $HR_nV_m$ Setting

The $HR_nV_m$ setting window allows the users to specify which $HR_nV_m$ analysis (choice of $m$ and $n$) to perform along with preprocessing options (Ectopic Beats) and frequency domain analysis options.

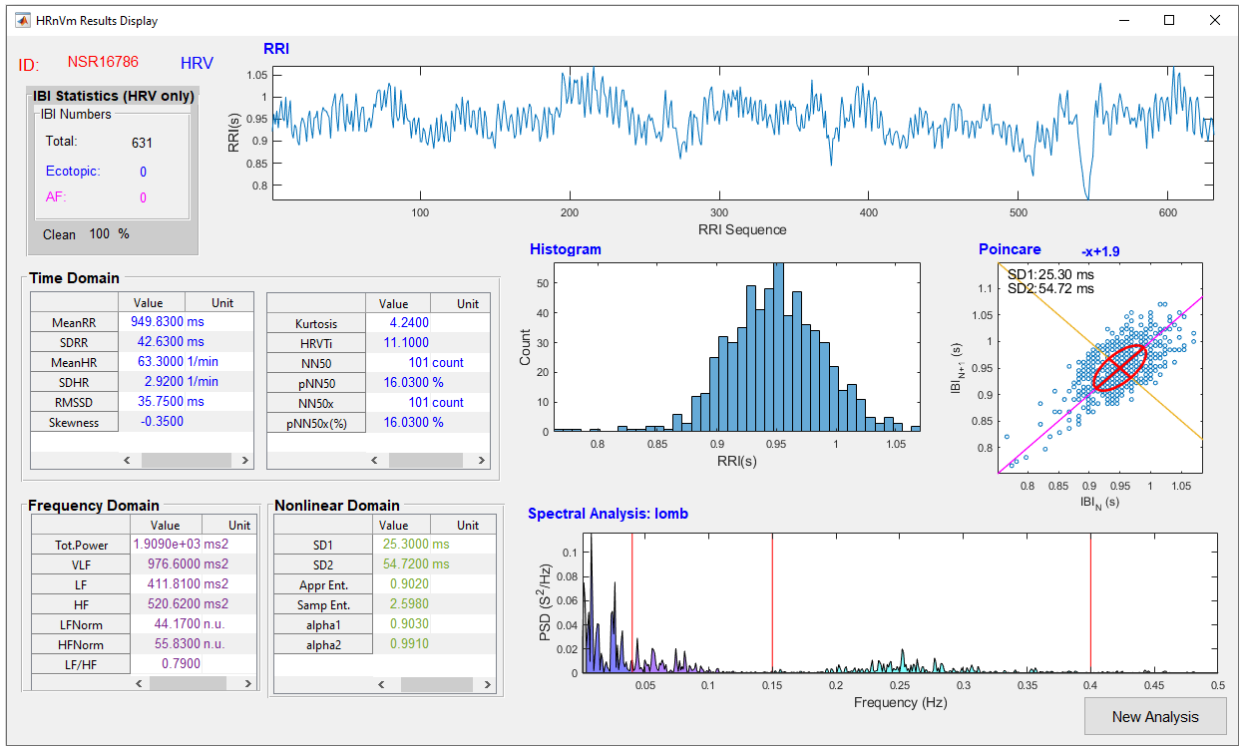

Figure 4 – $HR_nV_m$ Results Display

$HR_nV_m$ Results Display provides a comprehensive overview of the results from a single HRV or HRnV analysis

Table 1 – Summary of HRV/HRnV Metrics provided in HRnV-Calc

| Metrics | Units | Description |
|---|---|---|
| **Time Domain** | | |
| **Average RR** | ms | The mean of $RR_nI_m$ intervals |
| **SDRR** | ms | The standard deviation of $RR_nI_m$ intervals |
| **Average HR** | 1/min | The mean of heart rates |
| **SDHR** | 1/min | The standard deviation of heart rates |
| **RMSSD** | ms | Square root of the mean squared differences between successive RR intervals |
| **NN50(x)** | count | Numbers of $RR_nI_m$ intervals differ more than 50 ms from the previous intervals. For NN50x, the difference will be set to x times of 50 ms, where x = $n$ in the corresponding $HR_nV_m$ analysis |
| **pNN50(x)** | % | Percentage of NN50(x) intervals within the entire $RR_nI_m$ intervals |
| **RR Skewness** | - | The skewness of the $RR_nI_m$ intervals distribution |
| **RR Kurtosis** | - | The kurtosis of the $RR_nI_m$ intervals distribution |
| **RR Triangular Index** | - | The integral of the $RR_nI_m$ intervals histogram divided by the height of the histogram |
| **Frequency Domain** | | |
| **VLF, LF, and HF Peak frequencies** | Hz | The peak frequencies in the power spectral distribution (PSD) for VLF, LF, and HF bands |
| **VLF, LF, and HF Powers** | $ms^2$ | Absolute powers of VLF, LF, and HF bands |
| **VLF, LF, and HF Power Percentages** | % | The percentage for powers of VLF, LF, and HF bands within the overall spectrum |
| **LF and HF Normalized Powers** | n.u. | Normalized powers for LF and HF bands |
| **Total Power** | $ms^2$ | The overall power of the PSD |
| **LF/HF** | - | The ratio between the powers of LF and HF bands |
| **Nolinear Domain** | | |
| **Poincare SD1 and SD2** | ms | The width and length of the eclipse fitted in the Poincare plot |
| **App_Ent** | - | Approximate entropy |
| **Sam_Ent** | - | Sample entropy |
| **DFA $\alpha_1$ and $\alpha_2$** | - | Short-term and long-term fluctuations of detrended fluctuation analysis (DFA) |